# Molecular Dynamics Study of Conformations of Beta-Cyclodextrin and its Eight Derivatives in Four Different Solvents

Wasinee Khuntawee[1,2,3], Mikko Karttunen[4], and Jirasak Wong-ekkabut[1,2,3,*]

[1]Department of Physics, Faculty of Science, Kasetsart University, Bangkok 10900, Thailand

[2]Computational Biomodelling Laboratory for Agricultural Science and Technology (CBLAST), Faculty of Science, Kasetsart University, Bangkok 10900, Thailand

[3]Thailand Center of Excellence in Physics (ThEP Center), Commission on Higher Education, Bangkok 10400, Thailand

[4]Department of Chemistry and Department of Applied Mathematics, Western University, 1151 Richmond Street, London, Ontario N6A 5B7, Canada

*E-mail addresses: jirasak.w@ku.ac.th (J. Wong-ekkabut)

## Abstract

Understanding the atomic level interactions and the resulting structural characteristics is required for developing beta-cyclodextrin (βCD) derivatives for pharmaceutical and other applications. The effect of four different solvents on the structures of the native βCD and its hydrophilic (methylated βCD; MEβCD and hydroxypropyl βCD; HPβCD) and hydrophobic derivatives (ethylated βCD; ETβCD) were explored using molecular dynamics (MD) simulations and solvation free energy calculations. The native βCD, 2-MEβCD, 6-MEβCD, 2,6-DMβCD, 2,3,6-TMβCD, 6-HPβCD, 2,6-HPβCD and 2,6-ETβCD in non-polar solvents (cyclohexane; CHX and octane; OCT) were stably formed in symmetric cyclic cavity shape through their intramolecular hydrogen bonds. In contrast, βCDs in polar solvents (methanol; MeOH and water; WAT) exhibited large structural changes and fluctuations leading to significant deformations of their cavities. Hydrogen bonding with polar solvents was found to be one of the major contributors to this behavior: solvent-βCD hydrogen bonding strongly competes with intramolecular bonding leading to significant changes in structural stability of βCDs. The exception to this is the hydrophobic 2,6-ETβCD which retained its spherical cavity in all solvents. Based on this, it is proposed that 2,6-ETβCD can act as a sustained release drug carrier.

## 1. Introduction

Cyclodextrins (CDs) are cyclic oligosaccharides (ring-structured sugar compounds) commonly used in pharmaceutical and food industry for drug complexes and improved solubility, and as cholesterol removers, respectively.[1, 2] Perhaps the most famous application of CDs, however, is in the commercial odor remover Febreze in which CDs are used to capture "stinky" molecules.[3] Three of CDs, alpha CD (αCD), beta CD (βCD) and gamma CD (γCD), are naturally occurring and consist of α-(1,4) linked D-glucopyranose with six, seven or eight units, respectively. The general shape of all CDs is a truncated cone with hydrophilic outer surface and hydrophobic interior, Figure 1. Cyclodextrins' history, development and applications have been recently reviewed by Crini.[4]

In this work, we use molecular dynamics (MD) simulations and solvation free energy calculations to investigate conformational properties of the native βCD, four derivatives of methylated βCD (MEβCD), three derivatives of hydroxypropyl βCD (HPβCD), and one ethylated βCD (ETβCD) derivative. ETβCD is hydrophobic[5] while all the rest are hydrophilic. These systems were studied in four different solvents, cyclohexane (CHX), methanol (MeOH), octane (OCT) and water (WAT). The list of all systems is provided in Table 1. This focus is primarily motivated by the fact that in pharmaceutical applications, renal side effects have been reported for parenteral administration and suggested to be a result of poor water solubility.[6, 7] Despite previous studies regarding water solubility[8-11], complex stability[12, 13], bioavailability of βCD inclusion complexes[10, 14, 15], and improvements by substitutions of the hydroxyl groups with various functional groups, the molecular origin of these effects is not known.

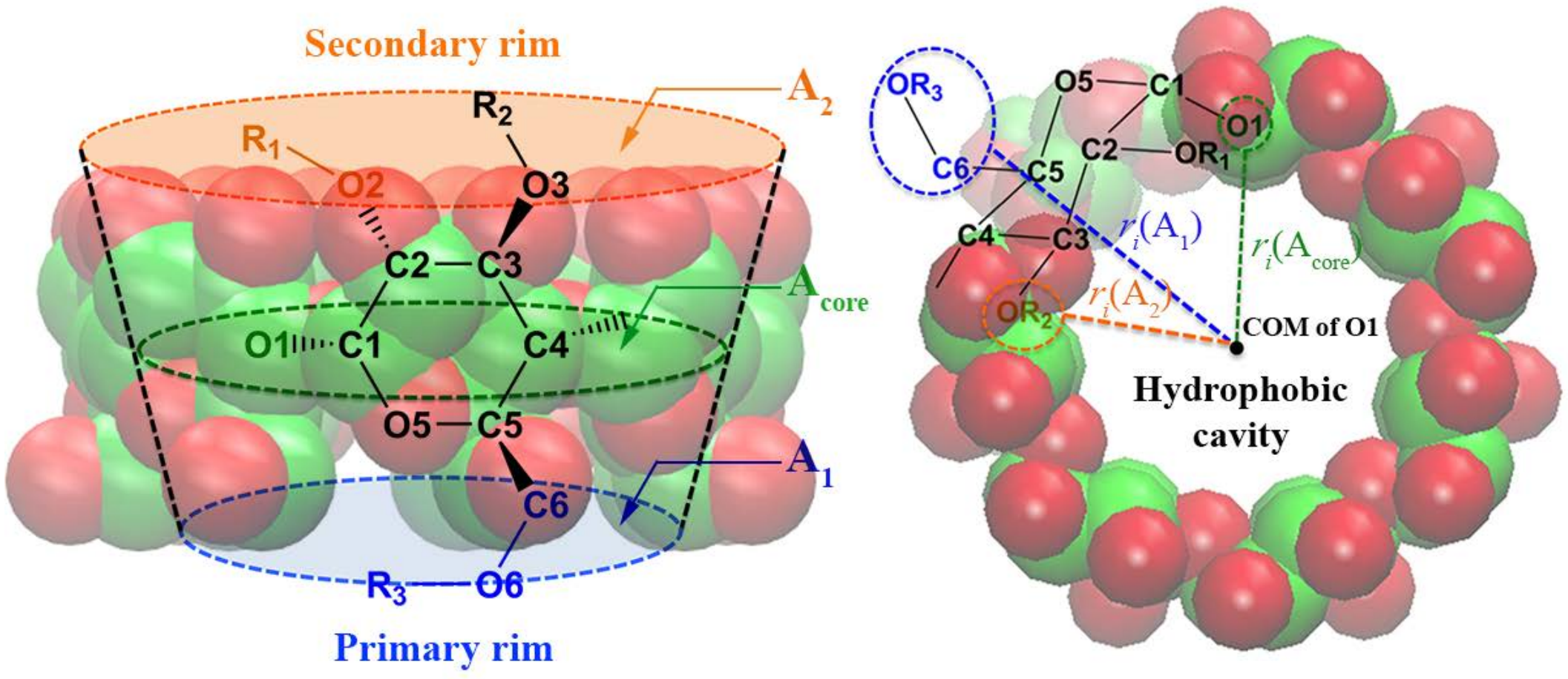

**Figure 1:** Left: Side-view of native βCD forming a truncated cone, showing the glucose subunit and atom name labeling. The rim at C6 is called the primary rim with the associated area $A_1$, while the opposite

rim, consisting of C2 and C3 is the secondary rim (area $A_2$). $A_{core}$ denotes the area at the center of the cavity. The R groups are varied for the βCD derivatives. $R_1$, $R_2$ and $R_3$ of all seven glucose subunits of the native βCD are hydrogen atoms. For the derivatives, $R_1$, $R_2$ or $R_3$ of are replaced by methyl ($-CH_3$), 2-hydroxypropyl ($-CH_2CH(OH)CH_3$) and ethyl ($-CH_2CH_3$) group, called methylated βCD, hydroxypropyl βCD and ethylated βCD, respectively. Right: Top-view of the βCD showing its hydrophobic cavity. The red and green spheres represent oxygen and carbon atoms, respectively. Hydrogen atoms are omitted for clarity.

Different functionalizations have been reported to alter structural, physicochemical and biological properties of βCDs.[16, 17] In addition, structural studies of several βCD types using X-ray diffraction and computer simulations have been conducted.[18-22] As a particular feature, Li et al.[23] found that the crystal structure of the native βCD is a truncated cone due to intramolecular hydrogen bonds (H-bonds) between the $R_1$ and $R_2$ groups of adjacent glucose subunits (Figure 1). Substitutions by methyl groups at $R_1$ and $R_3$, (Figure 1) called 2,6-dimethylated-β-CD (2,6-DMβCD; the numbers correspond to the numbering of the oxygen atom linking to those functional groups) narrowed the primary rim but the cavity still retained its cyclic shape due to intramolecular H-bonds.[24]

Structural characterization of βCD derivatives requires their synthesis which is rather difficult since substitutions at $R_1$, $R_2$ and $R_3$ compete with each other (Figure 1).[18] Computer simulations offer an alternative approach to study structure and conformational changes. For example, Yong et al.[22] used MD simulations to study the structural properties of HPβCD derivatives with varying numbers and positions of substituent groups in water. They found that structural changes in cavity shapes influence their interactions with guest ligands and the surrounding solvents and intra-molecular interactions. In another MD study, the rate constant for hydrogen bond breaking and reformation between βCDs and water around/inside cavity was observed to decrease for MEβCDs in comparison to the native βCD.[25]

Previous experiments have shown that water solubility of MEβCD and HPβCD can be enhanced by over 20-fold, compared to the native βCD.[26-29] In contrast, the solubility of ETβCD is three orders of magnitudes lower than that of the native βCD.[5, 30] This change in solubility most likely results from changes in intramolecular hydrogen bonding and hydrogen bonds with water. It has been shown, that toxicity depends on the number of functional groups and their positions[31] and that in drug delivery systems modified hydrophilicity due to substitutions results in different drug release profiles. In particular, hydrophilic βCD derivatives (see Table 1) can be used as immediate release transporters via increasing dissolution rate and adsorption of poorly water-soluble drugs, whereas hydrophobic derivatives act as sustained release drug carriers for water-soluble compounds.[32] Recent results also show that most poorly water-soluble drugs bind strongly to MEβCD and HPβCD derivatives which leads to a significant increase in solubility compared to free drugs as well as drugs complexed with the native βCD.[17] Several studies have also suggested that this improvement might be a result from changes in shape and solvent interactions of βCD derivatives[33, 34]; combination of CD complexation and co-solvation is one of the most promising techniques for improvement of drug solubility.[26, 34] Using alcohols (e.g. methanol, ethanol, etc.) as co-solvents, water solubility of guest ligands has been shown to increase.[35] The addition of non-polar solvent may also enhance the binding affinity of the guest ligand to the βCD's cavity.[36] Moreover, non-polar solvents have an important role in the purification process of CDs. In particular, cyclohexane helps to separate CDs from non-converted starch.[37] The precise molecular level mechanisms remain unresolved and thus detailed structural analyses are fundamental to understanding βCDs' properties. Resolving them is the aim of this paper.

## 2. Methodology

### 2.1) System preparation

Structural properties of the native βCD and its derivatives (MEβCD, HPβCD and ETβCD) were investigated in four different solvents (water, methanol, octane and cyclohexane) by atomistic MD simulations. The initial βCD configuration was taken from a previously relaxed βCD.[38] The starting structures of the derivatives were prepared from the native structure in which the hydrogen atoms of the hydroxyl groups at carbon positions 2-, 6-, 2,6- and 2,3,6- for all seven glucoses subunits were replaced by methyl groups, 2-hydroxypropyl groups and ethyl groups for βCD derivatives of MEβCD, HPβCD and ETβCD, respectively. The native βCD and eight different βCD derivatives are described in Table 1.

**Table 1:** Details of native βCD and its derivatives. The derivatives are classified into two main groups: 1) hydrophilic (MEβCDs and HPβCDs) and 2) hydrophobic (ETβCD) according to their water solubility with respect to the native βCD. The position and number of functional groups were varied for MEβCDs and HPβCDs. The number in the name of each βCD derivative corresponds to the number of oxygen atom connecting to the functional group R. The functional groups were fully substituted for all seven glucose subunits.

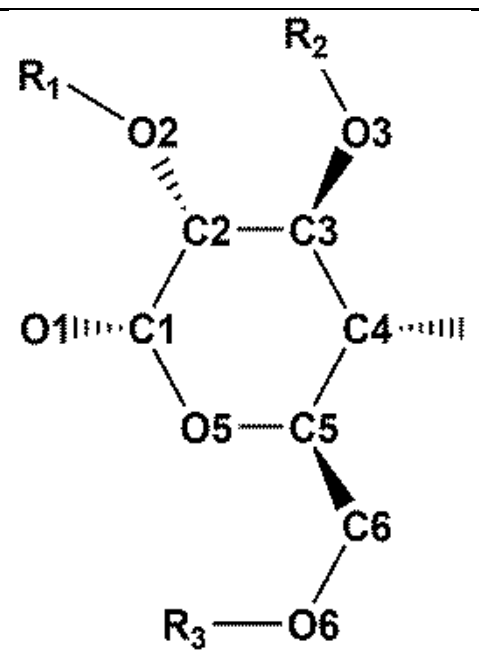

| Model | System | $R_1$ | $R_2$ | $R_3$ |
|---|---|---|---|---|
| a) Native βCD | | | | |
| 1 | βCD | -H | -H | -H |
| Hydrophilic βCDs | | | | |
| b) Methylated βCD derivatives | | | | |
| 2 | 2-MEβCD | $-CH_3$ | -H | -H |
| 3 | 6-MEβCD | -H | -H | $-CH_3$ |
| 4 | 2,6-DMβCD | $-CH_3$ | -H | $-CH_3$ |
| 5 | 2,3,6-TMβCD | $-CH_3$ | $-CH_3$ | $-CH_3$ |
| c) Hydroxypropyl βCD derivatives | | | | |
| 6 | 2-HPβCD | $-CH_2CH(OH)CH_3$ | -H | -H |
| 7 | 6-HPβCD | -H | -H | $-CH_2CH(OH)CH_3$ |
| 8 | 2,6-HPβCD | $-CH_2CH(OH)CH_3$ | -H | $-CH_2CH(OH)CH_3$ |
| Hydrophobic βCDs | | | | |
| d) Ethylated βCD derivative | | | | |
| 9 | 2,6-ETβCD | $-CH_2CH_3$ | -H | $-CH_2CH_3$ |

The GROMACS 5.1.1 package[39] was used to perform the MD simulations. The molecular models of native βCD, βCD derivatives, methanol, octane and cyclohexane were represented by the Gromos53a6 force field;[40, 41] we also tested the native βCD system with GLYCAM06 force field[42] and the results were similar. The partial charges and atom types of substituent groups in MβCD, HPβCD and ETβCD are shown in Figure S1. In simulations, the βCD in question was initially positioned at the center of the simulation box and fully solvated with 7000 single point charge (SPC) water molecules[43], 1728 methanol molecules, 1000 octane molecules or 2000 cyclohexane molecules depending on the solvent. The details of the simulated systems are shown in Table S1.

### 2.2) MD simulations

All initial structures were first energy minimized by using the steepest descent algorithm. This was followed by an MD simulation with a time step of 2 fs in the NPT (constant particle number, pressure and temperature) ensemble. The root mean square displacement (rmsd) of all atoms in the βCD molecules relative to their minimized structures was monitored and it was determined that the systems had reached equilibrium after 70 ns (Figure S2). Data collection for analysis started after that. The total simulation time for each of the systems was 100 ns. The Lennard-Jones and the real-space part of electrostatic interactions were cut-off at 1.0 nm. For long-range electrostatic interactions, the particle-mesh Ewald (PME) method[44-46] was used with the reciprocal-space interactions evaluated on a 0.12 nm grid with cubic interpolation of order four. The P-LINCS algorithm was used to constrain all bond lengths.[47] Isotropic pressure coupling was applied using the Berendsen algorithm[48] at 1 bar with a time constant of 3.0 ps and compressibility of $4.5 \times 10^{-5}$ $bar^{-1}$. The Parrinello-Donadio-Bussi velocity rescale thermostat algorithm was applied independently for βCD and water at 298 K.[49, 50] Periodic boundary conditions were applied in all directions. The above simulation protocol has been previously validated and used for several lipid and protein systems, for recent ones, see e.g. Refs.[51-54]. The Visual Molecular Dynamics (VMD) software was used for all molecular visualizations.[55]

## 3. Results and discussions

### 3.1) Structural changes in solvents

Structural changes from the energy-minimized structure were measured by the root mean square displacement (rmsd) for all atoms in the βCDs. Figures 2(a)-(i) show the rmsd distributions in different solvents. The averages rmsd are shown in Table S2. Fluctuations of the rmsd distributions can be discussed in terms of the full width at half maximum (FWHM) of the RMSD distributions, as shown in Table S3. The distributions were fitted to a Gaussian model and the FWHM values were calculated using $FWHM = 2\sqrt{ln4} \cdot \sigma$ where $\sigma$ is the standard deviation. In general, FWHM was lower in non-polar solvents than in polar ones with the following exceptions: 2,3,6-TMβCD in OCT, 2-HPβCD in OCT, 6-HPβCD in OCT and 2,6-ETβCD in CHX and OCT. Interestingly, the FWHM of the hydrophobic 2,6-ETβCD in polar solvents is smaller than in non-polar solvent. This is in contrast to the hydrophilic derivatives as 2,6-DMBCD and 2,6-HPBCD which have their substituent groups at the same positions.

The rmsd of the native βCD peaks around 0.11 nm in OCT, and around 0.12 nm, 0.20 nm and 0.26 nm in CHX, MeOH and WAT, respectively (Figure 2(a) and Table S2). The rmsd value of the native βCD in water is similar to the previous MD simulation using the same force field as us (Gromos53a6)[56]; the general βCD structural properties using Gromos53a6 are in agreement with X-ray scattering and simulations with other force fields[42, 57]. Compared to the native βCD in water, the rmsd peak position was about 23% smaller in MeOH. This tendency has been reported in previous simulations[21, 58], but the difference in their results was smaller by about 17%[58]. This may be due the difference in simulation times and solvation: our simulations were performed at higher solvation level and are an order of magnitude longer (10 vs 100 ns). In addition, as the rmsd time evolutions in Figure S2 show, structural changes can occur even at later times.

Compared to the native βCD, the peak of the rmsd distribution for βCD derivatives moves to higher values in all solvents except MEβCDs in water. For 2,6-ETβCD, the positions of the peaks were in the same range with the native βCD although their relative positions changed. For MEβCDs (Figures 2(b)-(e)), the lowest rmsd was found in non-polar solvents similar to the native βCD. The mono-substituted 2-MEβCD and 6-MEβCD showed large rmsd in OCT, while the rmsd of the di-substituted 2,6-DMβCD in CHX and OCT were similar. When the native βCD and MEβCDs were solvated in polar solvents, rmsd was increased. Moreover, structural change in WAT was less than in MeOH with the exception of 2-MEβCD. The fully substituted 2,3,6-TMβCD showed an increase in rmsd of 0.20-0.35 nm without any significant structural changes in different solvents.

In the case of the HPβCD derivatives, the long chain functional groups of 2-hydroxypropyl induced larger change in the rmsd compared to the other βCD types. The rmsds of the 2-HPβCD and 2,6-HPβCD were small in CHX and large in polar solvents, the largest in OCT for 2-HPβCD and in MeOH for 2,6-HPβCD. Similarly to 2,3,6-TMβCD, the structure of 6-HPβCD was relatively insensitive to the type of solvent. The peak of the rmsd of 6-HPβCD was in the range of 0.26-0.30 nm.

Finally, the hydrophobic 2,6-ETβCD was most unchanged in OCT. The structure underwent larger changes in CHX and polar solvents. The rmsds of the 2,6-ETβCD in WAT and MeOH were similar.

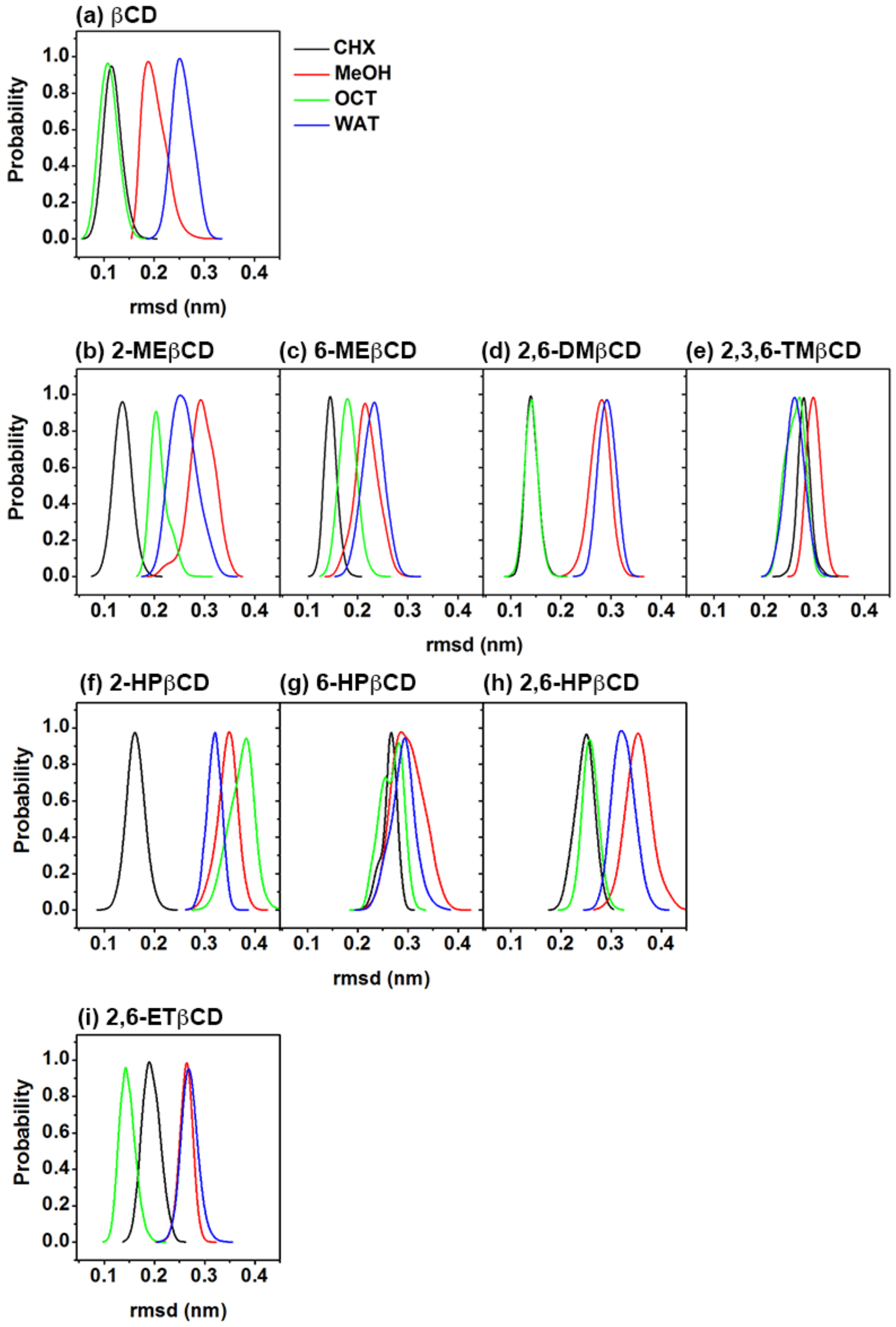

**Figure 2:** (a)-(i) The root mean square displacement (rmsd) distribution of the nine different βCDs in different solvents: CHX (black), MeOH (red), OCT (green) and WAT (blue). The polar solvents MeOH

and WAT induced larger structural changes in βCD derivatives with the exceptions of 2,3,6-TMβCD, 2-HPβCD and 6-HPβCD.

The root mean square fluctuations (rmsf) of atomic positions with respect to their initial coordinates were investigated (Figure 3(a)-(i)). The rmsf of each atom was averaged for the seven repeating glucose subunits, see atom labeling in Figure 1. The qualitative features of the rmsf profiles are the same in all systems with the exception of 2-HPβCD (Figure 3(f)) in which the middle peak is the highest one. In particular, the functional groups at the primary rim (at C6) exhibit more pronounced fluctuations compared to the functional groups at the secondary rim (at C2 and C3). In contrast, for 2-HPβCD (Figure 3(f)) large fluctuations were observed at the secondary rim at C2 functional groups.

Figure 3(a) shows that the native βCD exhibits less fluctuations in non-polar solvents. Compared to the βCD in WAT, the βCD in MeOH showed smaller fluctuations. This is in agreement with the previous simulations of Zhang et al.[58] Similarly, the MEβCD derivatives exhibit small fluctuations in non-polar solvents and increased rmsf in polar solvents. The difference between the rmsf in non-polar and polar solvents is shown for 6-MEβCD and 2,6-DMβCD (Figures 3(c) and 3(d)). For all MEβCD derivatives, 2-MEβCD displayed largest fluctuations. Among all the three HPβCD derivatives, 2-HPβCD has the smallest rmsf. The rmsf of 2-HPβCD has smallest fluctuations in WAT and fluctuations increase in CHX, MeOH and OCT.

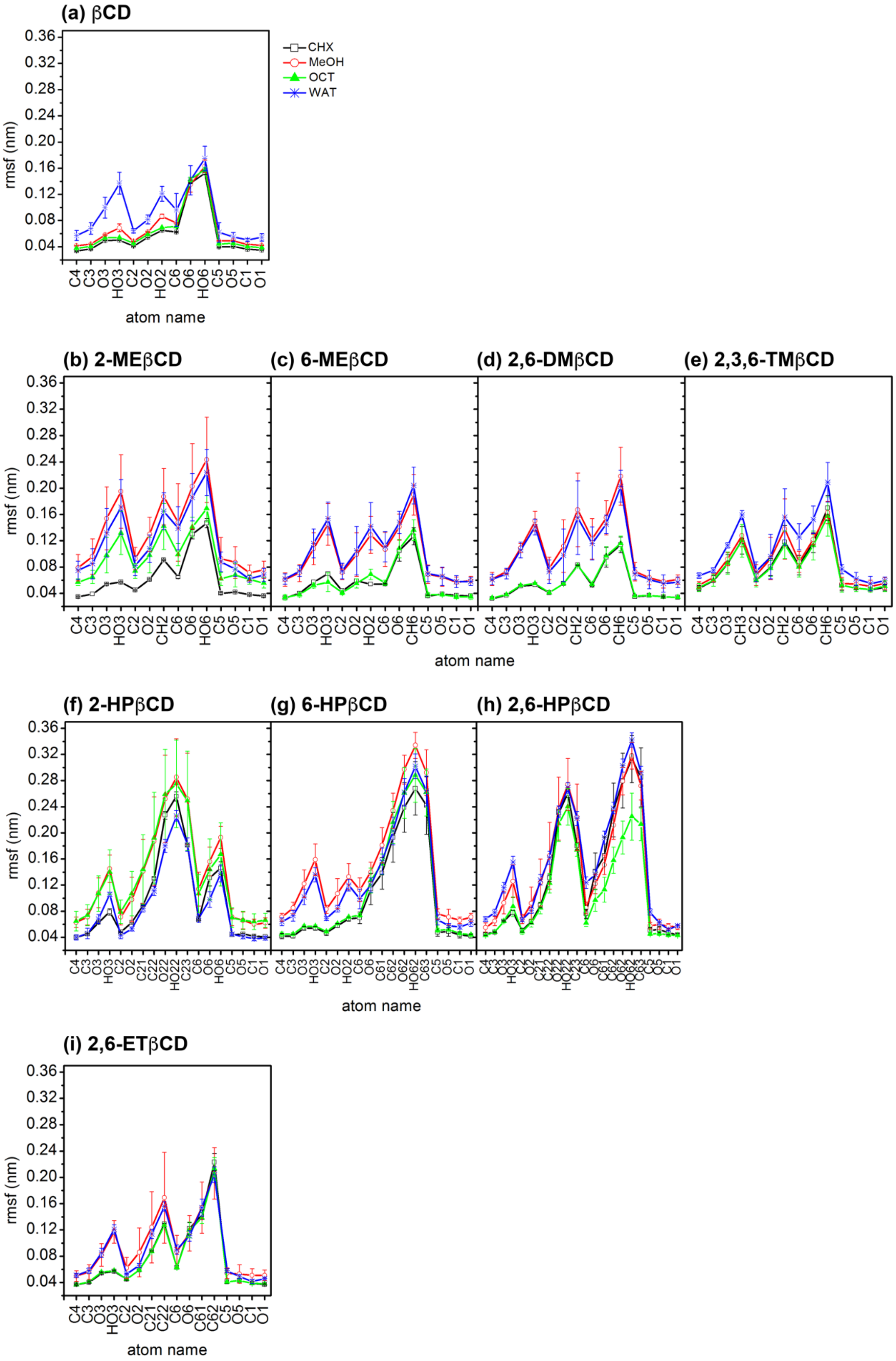

(a) βCD
CHX
MeOH
OCT
WAT
rmsf (nm)
0.36
0.32
0.28
0.24
0.20
0.16
0.12
0.08
0.04
C4 C3 O3 HO3 C2 O2 HO2 C6 O6 HO6 C5 O5 C1 O1
atom name
(b) 2-MEβCD
(c) 6-MEβCD
(d) 2,6-DMβCD
(e) 2,3,6-TMβCD
C4 C3 O3 HO3 C2 O2 CH2 C6 O6 HO6 C5 O5 C1 O1
C4 C3 O3 HO3 C2 O2 HO2 C6 O6 CH6 C5 O5 C1 O1
C4 C3 O3 HO3 C2 O2 CH2 C6 O6 CH6 C5 O5 C1 O1
C4 C3 O3 CH3 C2 O2 CH2 C6 O6 CH6 C5 O5 C1 O1
atom name
(f) 2-HPβCD
(g) 6-HPβCD
(h) 2,6-HPβCD
C4 C3 O3 HO3 C2 O2 C21 C22 O22 HO22 C23 C6 O6 HO6 C5 O5 C1 O1
C4 C3 O3 HO3 C2 O2 HO2 C6 O6 C61 C62 O62 HO62 C63 C5 O5 C1 O1
atom name
(i) 2,6-ETβCD
C4 C3 O3 HO3 C2 O2 C21 C22 C6 O6 C61 C62 C5 O5 C1 O1
atom name

**Figure 3:** (a)-(i) The averages of root mean square fluctuations (rmsf) of the nine different βCDs in different solvents; CHX (black), MeOH (red), OCT (green) and WAT (blue), the higher fluctuation of βCDs structure was mostly found in polar solvents (WAT and MeOH), compared to non-polar solvents (OCT and CHX). Only the 2-HPβCD was more stable in WAT.

### 3.2) Hydrogen bonding

In addition, the number of intramolecular hydrogen bonds (H-bond) may have a significant impact on the structural stability of βCDs.[59-61] The number of hydrogen bonds between the -OR$_1$ group and the -OR$_2$ group of the adjacent glucose subunits was monitored in each of the cases. As detailed in Figure 4(a), the native βCD formed on average about 7 intramolecular H-bonds in both of the non-polar solvents, while only few hydrogen bonds were found in polar solvents. Being solvated in CHX, the number of adjacent H-bonds of the -OR$_1$ and -OR$_2$ groups for the native βCD were the same as for the βCD derivatives. There is an exception, however: for 2-HPβCD and 2,6-HPβCD, the number of adjacent H-bonds for the -OR$_1$ and -OR$_2$ groups are higher than for the native βCD. In OCT, the number of adjacent H-bonds was in the same range as in CHX, except for 2-MEβCD and 2-HPβCD. These results correspond to the comparison of structural changes in CHX and OCT. In polar solvents, the adjacent H-bond for the native βCD was smaller than for the βCD derivatives, especially in MeOH. The loss of intramolecular H-bonds of βCDs resulted from increased intermolecular H-bonding between the βCDs and polar solvents (Table 2). Similar effects were seen in all βCD derivatives albeit with some interesting characteristics that will be discussed in the next section in connection with the shape analysis.

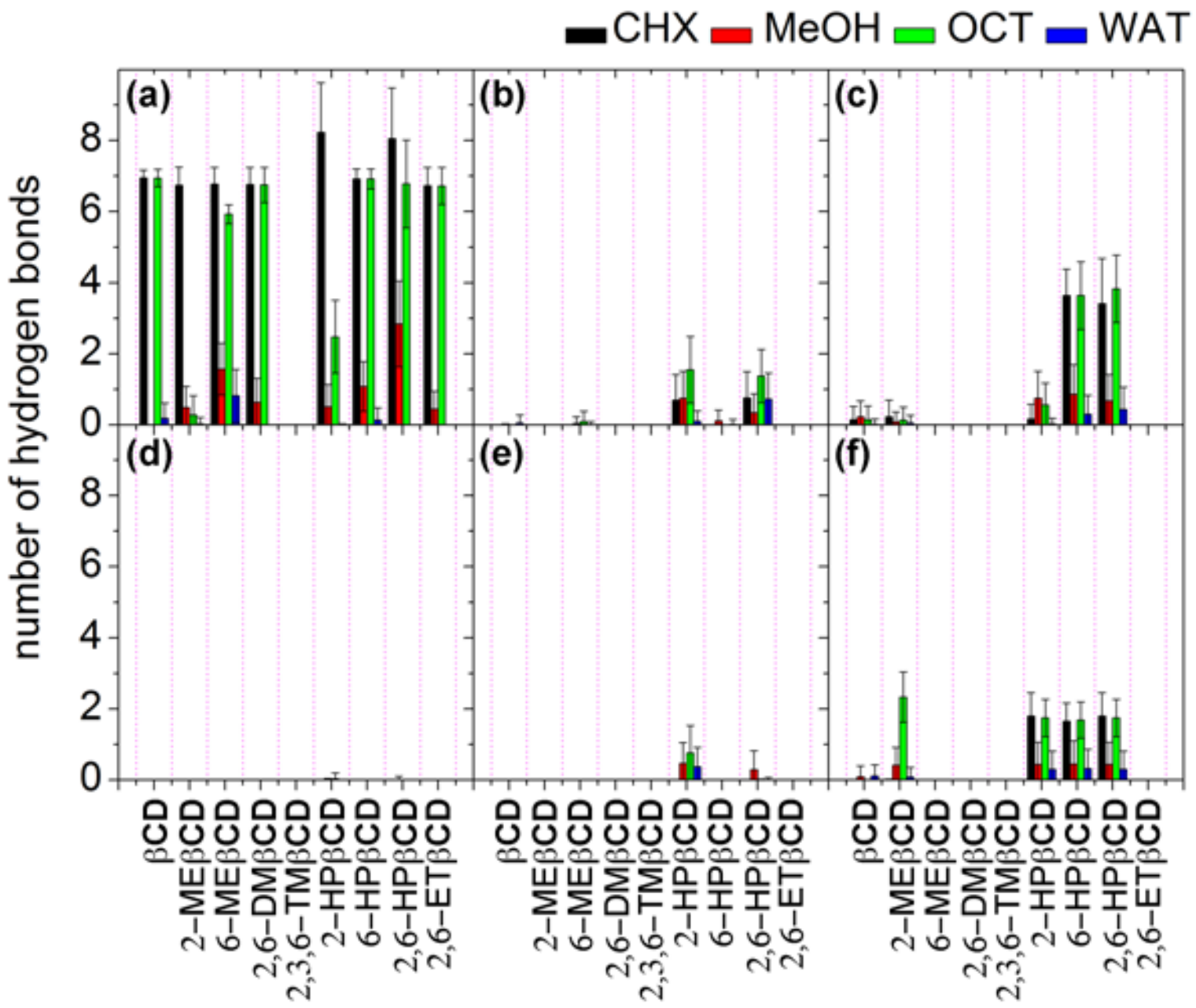

**Figure 4**: (a)-(c) intramolecular hydrogen bonds between the adjacent glucose subunits (shown in Figure 1): (a) the -$OR_1$ and -$OR_2$ groups, (b) the -$OR_1$ and -$OR_1$ groups, and (c) the -$OR_3$ and -$OR_3$ groups. Figures (d)-(f) show intramolecular hydrogen bonds between the non-adjacent glucose subunits: (d) the-$OR_1$ and -$OR_2$ groups, (e) the -$OR_1$ and -$OR_1$ groups, and (f) the -$OR_3$ and -$OR_3$ groups. There are no hydrogen donors or acceptors in 2,3,6-TMβCD.

**Table 2:** The number of hydrogen bonds between the βCDs and polar solvents. Error is given as standard deviation.

| System | Average numbers of H-bonds | |
|---|---|---|
| | MeOH | WAT |
| βCD | 32.8±3.0 | 42.3±3.1 |
| 2-MEβCD | 22.3±2.8 | 34.0±2.9 |
| 6-MEβCD | 21.9±2.8 | 32.3±3.0 |
| 2,6-DMβCD | 15.2±2.3 | 25.0±2.5 |
| 2,3,6-TMβCD | 6.9±1.8 | 15.2±2.4 |
| 2-HPβCD | 29.1±3.3 | 45.1±3.3 |
| 6-HPβCD | 30.9±3.5 | 44.2±3.4 |
| 2,6-HPβCD | 30.1±3.4 | 47.0±3.8 |
| 2,6-ETβCD | 14.8±2.3 | 23.7±2.5 |

Our results suggest that non-polar solvents (CHX and OCT) may stabilize the structures for most of the βCDs except for 2-HPβCD in OCT. The deformation of 2-HPβCD in OCT could be found because some substituent flipped toward inside the cavity and interacted with their non-neighbor substituents (Figure S3). Moreover, the inclusion of the OCT inside the 2-HPβCD's cavity was not found, while the CHX could be bound to the cavity (Figure S3). The inclusion complex of non-polar solvents inside the βCDs' cavity may also play role in the βCDs structure stabilization. Interestingly, 2,6-ETβCD shows lesser structural changes in OCT as compared to the other βCD derivatives. Molecules of polar solvents, water and methanol, may be present inside the cavity interior as shown in Figures S4 and S5. Water molecules present inside the native βCD cavity were found similarly to the X-ray crystal structures [62, 63]. For βCD derivatives, the number of water molecules inside the cavity of difunctionalized βCD derivatives was significantly higher than in monofunctionalized βCDs. A few methanol molecules were observed inside cavity, except for 2-MEβCD and 2-HPβCD. No methanol molecules were present inside the deformed cavity of those βCDs. Molecules of the polar solvents were located at the hydrogen acceptors and hydrogen donors of the βCDs, that is, not inside the cavity. Hydrogen bonds with polar solvents were formed resulting in structural deformation of βCDs. Polar solvents caused higher fluctuations in βCDs' structures, especially for the native βCD and the MEβCD derivatives. The structural changes of βCDs as well as their shapes may be factors in altering guest ligands' binding to the cavity interior. The influence of solvents on the βCDs' shapes will be discussed in the next section.

**3.3) Shape of βCDs**

The radius of gyration ( $R_g$ ) and asphericity ($b$) were examined to describe sizes and shapes. The three principal moments ($\lambda_1$, $\lambda_2$ and $\lambda_3$ where $\lambda_1^2 \geq \lambda_2^2 \geq \lambda_3^2$) following the common ordering convention) of the $R_g$ tensor were measured. $R_g$ can be given in terms of the principal moments as $R_g = \sqrt{\lambda_1^2 + \lambda_2^2 + \lambda_3^2}$ and asphericity as $b = \lambda_1 - \frac{1}{2}(\lambda_2 + \lambda_3)$. For a spherically symmetric object $b = 0$.

To explore the local structural properties, the areas ($A$) of core structure (Figure 1) at each rim were calculated using

$$A = \frac{\pi}{7}\sum_{i=1}^{7} r_i^2, \tag{1}$$

where $r_i$ is the distance between the βCD's center and the group of atoms of interest in glucose subunit *i*. The βCD's center was determined as the center of mass (COM) of all O1 atoms. The groups of interest are: *1)* O1 atoms, *2)* C6···O6···$R_3$ groups, and *3)* O2···$R_1$ groups in glucose subunits. They were used to represent the cavity area at the core structure ($A_{core}$), the primary rim ($A_1$) and the secondary rim ($A_2$), respectively. The definitions of areas are shown in Figure 1.

The averages of $R_g$ and $b$ are shown in Table 3. The time evolutions of $R_g$ and its three principal components ($\lambda_1$ , $\lambda_2$ and $\lambda_3$) are plotted in Figure S6. Additionally, snapshots from the final configurations at *t*=100 ns are shown in Figure 5. As compared to the native βCD, the $R_g$ are in the same range (0.61-0.65 nm) for the MEβCD and increased for HPβCD and ETβCD. The increase of $R_g$ in water is in quantitative agreement with previous simulations of βCD and HPβCD.[22] For the different solvents, the $R_g$s do not show significant differences. Circularity can be examined by using the three principal components; when two of the principal components are equal, the planar structure is circular, the smallest value is in the direction along the cylindrical axis. Their time evolutions (Figure S6) suggest that the native βCD is very close to circular with the exception of water solution where the two largest principal components attain different values after about 10 ns. Regarding all derivatives, the highest degree of circularity is observed in CHX. As Figure S6 also shows, it is clear that long simulations times are needed to capture structural changes. In addition, in polar solvents (MeOH and WAT) the native βCD showed higher asphericity than in non-polar solvents by 22% and 56%, respectively (in Table 2). For the MEβCD derivatives in non-polar solvents, the cavity mostly formed a circular shape with the exception of 2,3,6-TMβCD. The 12-38% difference in $\lambda_1$ and $\lambda_2$ for 2,3,6-TMβCD in all solvents indicates the cavity to be ellipsoidal. This is an agreement with X-ray studies.[19] Compared to solvation in CHX, solvation in MeOH showed increasing asphericity by 62%, 50% and 68% for 2-MEβCD, 6-MEβCD and 2,6-DMβCD, respectively. In the case of the HPβCD derivatives, most of the HPβCDs in non-polar solvents had an approximately spherical cavity. In contrast, the HPβCD cavity in polar solvents was elliptical: large differences between $\lambda_1$ and $\lambda_2$ values, in the range of 8-37%, were found, especially for 2-HPβCD in MeOH (37%) and OCT (32%). In the case of the di-substituted 2,6-HPβCD, $\lambda_1$ and $\lambda_2$ showed no significant dependence on the type of solvent. However, $\lambda_3$ increased to be in the same range with $\lambda_1$ and $\lambda_2$ especially when the 2,6-HPβCD was solvated by WAT. The HPβCD derivatives with substitutions at both 2- and 6-positions were more spherical than the substitutions at only one of those positions. This is in agreement with the simulations of HPβCD derivatives in water.[22] Most of the HPβCDs in CHX were more spherical than in the other solvents; the 2,6-HPβCD in WAT has the lowest asphericity. The spherical shape was highly deformed in OCT and in MeOH in case of 2-HPβCD and the change occurred after a significant time (Figure S6). Finally, in case of the 2,6-ETβCD, $\lambda_1$ and $\lambda_2$ fluctuated in the same range independent of the type of solvent. It indicates that the circular cavity of 2,6-ETβCD was maintained in all solvents. The $\lambda_3$ of the 2,6-ETβCD in CHX and OCT were similar. By comparing in CHX, the decrease of $\lambda_3$ was found by 19% and 10% when the 2,6-ETβCD was solvated by MeOH and WAT, respectively. Interestingly, the 2,6-ETβCD remained spherical in all solvents ($b$ ~ 0.08-0.09).

**Table 3:** The effect of solvent on the radius of gyration ($R_g$) and asphericity ($b$). Most of the βCDs are larger and more spherical in non-polar solvents than in polar solvents. Errors are given in terms of standard deviation. The errors in $R_g$ and $b$ are less than 0.01 and 0.03, respectively.

| System | Radius of gyration; $R_g$(nm) | | | | Asphericity; $b$ | | | |
|---|---|---|---|---|---|---|---|---|
| | **CHX** | **MeOH** | **OCT** | **WAT** | **CHX** | **MeOH** | **OCT** | **WAT** |
| βCD | 0.62±0.01 | 0.62±0.01 | 0.62±0.01 | 0.60±0.01 | 0.09±0.01 | 0.11±0.01 | 0.09±0.01 | 0.14±0.02 |
| 2-MEβCD | 0.64±0.01 | 0.61±0.01 | 0.60±0.01 | 0.62±0.01 | 0.09±0.01 | 0.15±0.02 | 0.10±0.01 | 0.13±0.02 |
| 6-MEβCD | 0.62±0.01 | 0.62±0.01 | 0.61±0.01 | 0.61±0.01 | 0.08±0.01 | 0.12±0.02 | 0.09±0.01 | 0.12±0.02 |
| 2,6-DMβCD | 0.64±0.01 | 0.62±0.01 | 0.64±0.01 | 0.62±0.01 | 0.08±0.01 | 0.13±0.02 | 0.08±0.01 | 0.10±0.02 |
| TMβCD | 0.63±0.01 | 0.63±0.01 | 0.63±0.01 | 0.65±0.01 | 0.13±0.01 | 0.14±0.01 | 0.15±0.01 | 0.11±0.01 |
| 2-HPβCD | 0.70±0.01 | 0.66±0.01 | 0.65±0.01 | 0.66±0.01 | 0.10±0.01 | 0.16±0.02 | 0.14±0.02 | 0.10±0.01 |
| 6-HPβCD | 0.64±0.01 | 0.66±0.01 | 0.64±0.01 | 0.65±0.01 | 0.09±0.02 | 0.12±0.03 | 0.09±0.02 | 0.12±0.03 |
| 2,6-HPβCD | 0.72±0.01 | 0.72±0.01 | 0.71±0.01 | 0.69±0.01 | 0.08±0.02 | 0.11±0.02 | 0.08±0.02 | 0.06±0.01 |
| 2,6-ETβCD | 0.67±0.01 | 0.65±0.01 | 0.67±0.01 | 0.64±0.01 | 0.08±0.01 | 0.09±0.01 | 0.08±0.01 | 0.09±0.01 |

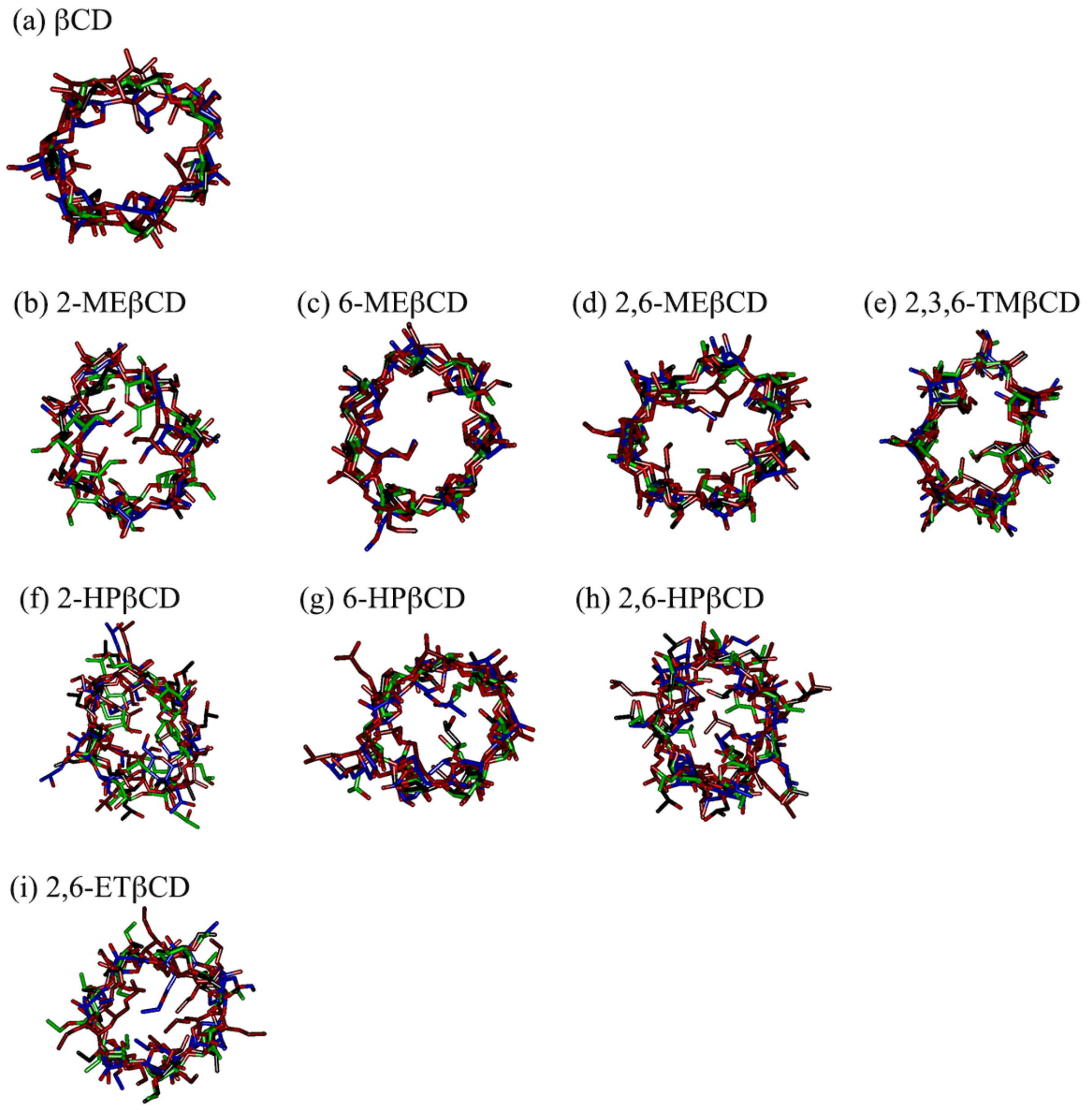

**Figure 5:** Superposition of the last MD snapshot of each βCD type in different solvents, the native βCD and βCD derivatives in CHX, MeOH, OCT and WAT are represented by the black-, red-, green- and blue-stick, respectively. The cyclic drastic deformation was found in polar solvents, especially for the 2,3,6-TMβCD, 2-HPβCD and 2,6-HPβCD.

The cavity areas of the core structure ($A_{core}$), the primary rim ($A_1$) and the secondary rim ($A_2$) are shown in Figures 6(a)-(i), the definitions for the areas are provided in Figure 1. The results show that for all βCD types, $A_{core}$ does not depend significantly on solvent type. At the rims, the area $A_2$ was more influenced by the solvent type than $A_1$. For the native βCD, the $A_2$was larger than $A_1$ in non-polar solvents. In contrast, the area at the primary rim was larger than at the secondary rim in polar solvents. Solvation of the native βCD in MeOH leads to a narrow secondary rim.

For the MEβCD derivatives (Figure 6(b)-(e)), cavity sizes show dependence on functionalization. In non-polar solvents, $A_2$ of the 2-MEβCD and 2,6-DMβCD increased to ~1.8 $nm^2$ whereas the native βCD and the rest of the MEβCD derivatives had $A_2$ ~ 1.4 $nm^2$. Relatively open secondary rims were found in non-polar solvents for 2-MEβCD and 2,6-DMβCD, compared to their primary rims. In polar solvents, however, $A_1$ and $A_2$ were similar for most of the MEβCD derivatives.

Figure 6(e) shows that there is no solvent effect on the TMβCD cavity size. In case of the HPβCD derivatives (Figure 6(f)-(h)), the area of substituted rim was larger than the rim without functional groups. Functionalization with long chain of 2-hydroxypropyl resulted in large areas compared to the native βCD and the other βCD derivatives. The areas at all parts of 2-HPβCD did not show any dependence on the type of solvent and 2-HPβCD had its secondary rim more open than the primary. In contrast, the primary rim of 6-HPβCD was more opened in all solvents. When both rims had substitutions, $A_2$ of 2,6-HPβCD was larger than $A_1$ in most of the solvents. The only exception was water. For 2,6-ETβCD, the secondary rim was larger in non-polar solvents and $A_1$ was equal to $A_2$ in polar solvents.

Shape analysis shows that βCDs in non-polar solvents have mostly spherical cavities whereas cavity deformations were found in polar solvents. The type of functionalization also had an influence on the cavity shape. Substitution at only one rim showed less circularity compared to the MEβCD and HPβCD with functional groups on their both rims. However, no significant changes in the area at the core (Figure 1) for different functional groups were observed. However, among the three functional groups, substitution with hydroxypropyl showed slightly larger area at the rims, especially at the rim(s) with the substituent.

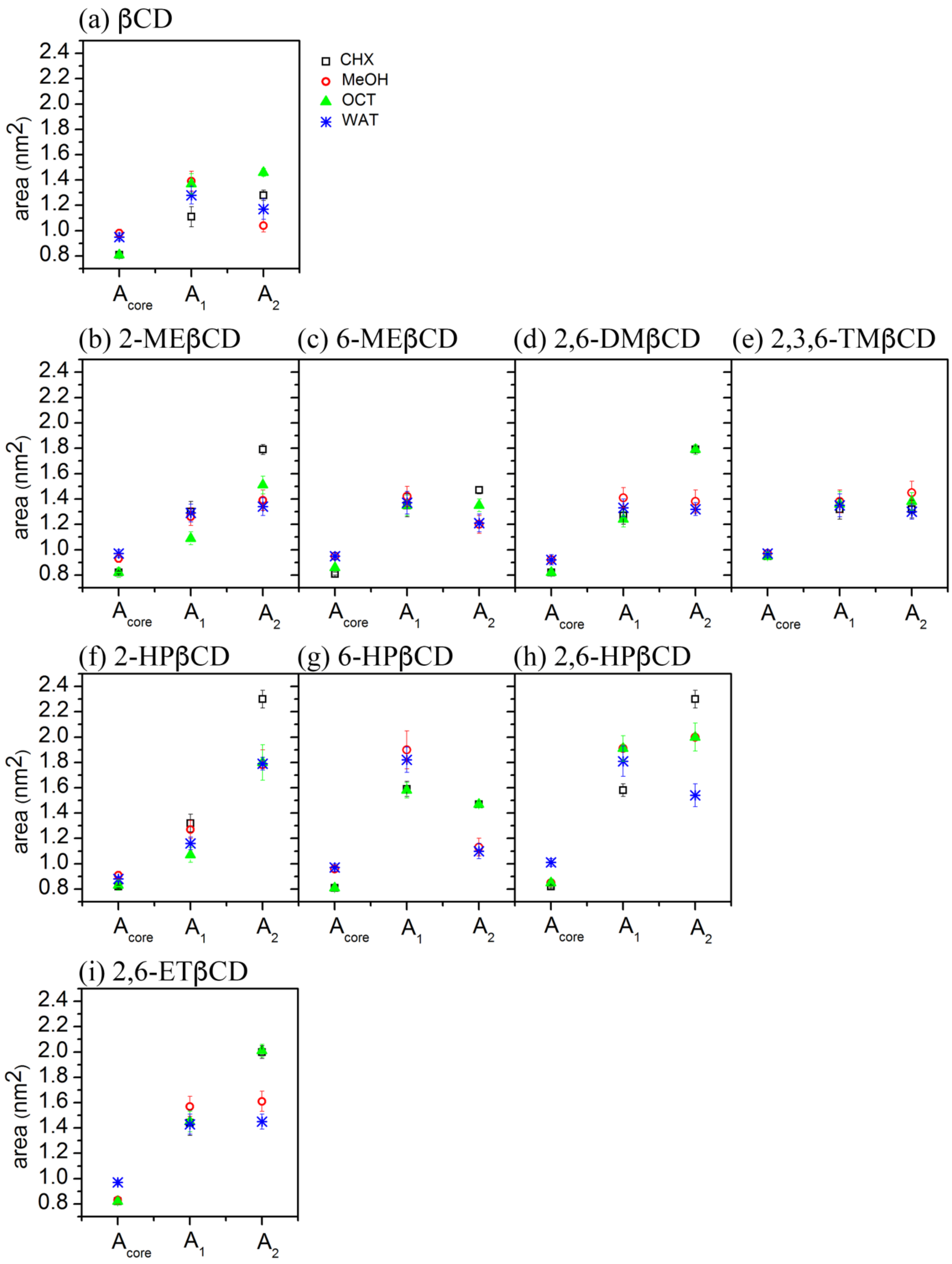

**Figure 6:** The area (Figure 1) at the core ($A_{core}$), the primary rim ($A_1$) and the secondary rim ($A_2$) in the presence of different solvents. There was no significant change in $A_{core}$ with different solvents or functional groups. $A_2$ was mostly larger than $A_1$ in non-polar solvents.

Before leaving this section, we discuss the relation between principal components and intramolecular hydrogen bonding. The case of native βCD has already been addressed above and so we focus on the βCD derivatives. Comparison of the time evolution of the principal moments (Figure S6) and the number of hydrogen bonds between the different glucose subunits (Figure 4) reveals the stabilizing influence of H-bonds between the adjacent -OR1 and -OR2 groups (Figure S7(a)) on the secondary rim, and the destabilizing effect of the H-bonds between the -OR3 groups (Figure 4c,f). In particular, when H-bonds between the -OR3 groups exist, fluctuations in the principal moments (Figure S6) become very pronounced. That is exemplified by the behavior of all HPβCDs. 2,3,6-TMβCD is another special case as it does not have any intramolecular H-bonds and it also shows large fluctuations. Side and top views of few of the cases are shown in Figure S7.

### 3.4) Solvation free energies

Solvation free energies ($G_{solvation}$) were estimated using the Molecular Mechanic/Poisson-Boltzmann Surface Area (MM/PBSA) method.[64] $G_{solvation}$ is the free energy difference between the solute in solvent and vacuum. It is composed of contributions due to electrostatic ($G_{polar}$) and non-electrostatic ($G_{non\text{-}polar}$) terms. $G_{polar}$ is estimated using a Poisson-Boltzmann model. The dielectric constant of the βCDs molecule was set to be equal to one.[65] The dielectric constants of the solvents were extracted from experiments.[66] The non-polar contribution depends on the βCD's geometry. The MM/PBSA calculation was performed at the rate of every 1 ns for the last 30 ns of MD trajectory. We would like to mention issues. First, MM/PBSA is a so-called end-point method, that is, only the free energy difference between two states is considered. Thus, it does not take entropic contributions fully into account. A recent review of free energy methods discussing MM/PBSA and alternatives is provided by Hansen and van Gunsteren.[67] The second issue is that solubility is not determined by solvation free energy alone. To properly account for solvation, the free energy of the solid phase should also be taken into account. A recent review is provided by Skyner et al..[68]

The average $G_{solvation}$, and the components $G_{polar}$ and $G_{non\text{-}polar}$ are shown in Figure 7. The main contribution to the free energy was observed to be always due to the polar interactions. The non-polar contribution in all cases constituted less than 30% of the total solvation free energy. The lowest non-polar contribution in water was found for the native βCD, followed by 6-MEβCD, 2-MEβCD, 2,6-DMβCD, 6-HPβCD, 2-HPβCD, TMβCD, 2,6-ETβCD and 2,6-HPβCD, respectively. The results in Figure 7 suggest that all βCDs favor polar solvents. In bulk water, the order for $G_{solvation}$ was TMβCD > 2,6-ETβCD > 2,6-DMβCD > 6-MEβCD > 2-MEβCD > βCD > 2-HPβCD ~ 6-HPβCD > 2,6-HPβCD. This order correlates well with hydrogen bonding (Table 2). The solvation free energies are qualitative agreement with experiments using the HPβCD and ETβCD derivatives.[26, 30]

In MeOH, $G_{solvation}$ is higher compared to water. The same trend as in water was observed with one exception: there was no significant difference in $G_{solvation}$ between the HPβCD derivatives. In non-polar solvents, $G_{solvation}$ was observed to be about five times higher than in polar solvents. TMβCD has the highest $G_{solvation}$ in CHX, followed by the 2,6-ETβCD, 2,6-DMβCD, 6-MEβCD, 6-HPβCD, 2,6-HPβCD, 2-MEβCD, βCD and 2-HPβCD. The order is the same in OCT.

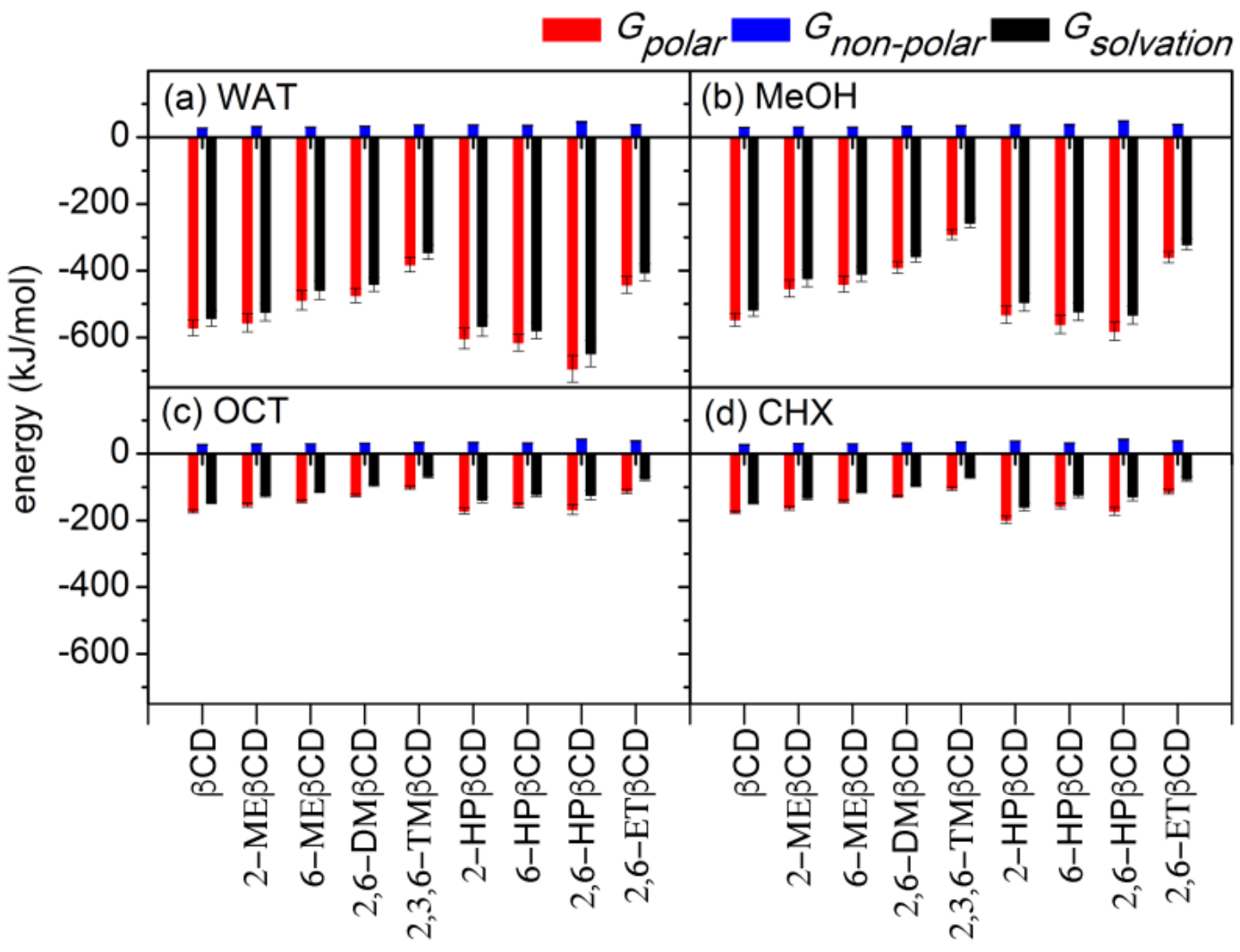

**Figure 7:** Solvation free energy, $G_{solvation}$ (black), and contributions from polar (red) and non-polar (blue) interactions. For the same type of βCD, $G_{solvation}$ was always lowest in WAT, followed by MeOH and non-polar solvents, respectively.

## 4. Conclusions

In the present work, conformational properties of the native βCD and eight of its derivatives, both hydrophilic and hydrophobic types, in four different solvents were investigated by MD simulations. Our results show that the polar solvents have strong influence on the structural stability of βCDs: intramolecular hydrogen bonds were lost, resulting in deformation of the βCDs' ring and decreased structural stability. An interesting exception to this behavior was solvation in octane which induced less stability and significant changes in the 2-HPβCD structure.

Interestingly, the hydrophobic 2,6-ETβCD structure showed high rigidity and the spherical shape of the cavity remained intact in all solvents. We propose that this high stability, which correlates well with its high ligand-binding affinity, may be the reason why 2,6-ETβCD can act as a sustained release drug carrier. The effect of polar solvents on the other βCD types was very different and both the positions and number of functional groups influenced their shape. In the case of di-substitution at C2 and C6, MEβCDs and HPβCDs had spherical cavity, while the mono-substituted ones had elliptical cavities. In addition, in non-polar solvents the secondary rim (Figure 1) remained relatively open while it was narrowed in polar solvents. The long chain of 2-hydroxypropyl functional groups of the HPβCD derivatives resulted in larger areas (Figure 1), especially at the substituted rim. The MM/PBSA calculations showed that the solvation free energy of each βCDs type was different depending on their chemical functional groups and the numbers of the substituent groups. All βCDs preferred solvation by polar solvents. In general, the atomistic details of the conformations in various solvents are highly useful for the selection of the appropriate βCDs in pharmaceutical applications and other applications, and in the development in drug delivery systems.

## ASSOCIATED CONTENT

### Supporting Information

The Supporting Information is available free of charge.

### Notes

The authors declare no competing financial interest.

### Acknowledgements

This work was financially supported by Kasetsart University Research and Development Institute (KURDI) and Faculty of Science at Kasetsart University (JW), and Natural Sciences and Engineering Research Council of Canada (MK).